# Proton FLASH irradiation platform for small animal setup at Chang Gung Memorial Hospital

Tung-Yuan Hsiao[1], Lu-Kai Wang [2,3], Tzung-Yuang Chen[1] ,Ching-Fang Yu[2,4], Pan, Cheng-Ya[4], Chun-Chieh Wang[2,4,5,6], Chien-Yu Lin[2,4], I-Chun Cho[7], Huan Niu[1], and Chien-Hsu Chen[1,*]

[1] *Accelerator Laboratory, Nuclear Science and Technology Development Center, National Tsing Hua University, Hsinchu, Taiwan*
[2] Radiation Research Core Laboratory, Institute for Radiological Research, Chang Gung University / Chang Gung Memorial Hospital, Linkou Branch, Taoyuan, Taiwan
[3] Department of Life Sciences, National Science and Technology Council, Taipei, Taiwan
[4] Department of Radiation Oncology, Chang Gung Memorial Hospital Linkou Branch, Taoyuan, Taiwan
[5] Radiation Biology Research Center, Institute for Radiological Research, Chang Gung University, Taoyuan, Taiwan
[6] Department of Medical Imaging and Radiological Sciences, Chang Gung University, Taoyuan, Taiwan
[7] Medical Physics Research Center, Institute for Radiological Research, Chang Gung University/Chang Gung Memorial Hospital Linkou Branch, Taoyuan Taiwan

[*] Author to whom correspondence should be addressed: akiracc@gmail.com, achchen@mx.nthu.edu.tw

ABSTRACT

Background :
Proton flash therapy is an emergency research topic in radiation therapy since the Varian announced the promising results from the first in human clinical trial of Flash therapy recently. However, it still needs a lot of researches on this topic, not only to understand the mechanism of the radiobiological effects but also to develop an appropriate dose monitoring system.
Purpose : In this study we setup an experimental station for small animal proton Flash irradiation in a clinical machine. The dose monitoring system is able to provide real-time irradiation dose and irradiation time structure.

Methods : The dose monitoring system includes homebrewed transmission ionization chamber (TIC), plastic scintillator based beam position monitor, and Poor Man Faraday Cup (FC). Both TIC and FC are equipped with a homebrewed fast reading current integral electronics device. The imaging guidance system comprises a moveable CT, laser, as well as attaching a bead on the body surface of the mouse can accurately guide the testing small animal in position.

Results : The dose monitoring system can provide the time structure of delivered dose rate within 1 ms time resolution. Experimental testing results show that the highest dose in one pulse of 230 MeV proton that can be delivered to the target is about 20 Gy during 199 ms pulse period at 100 Gy/s dose rate.

Conclusion : A proton research irradiation platform dedicated for studying small animal Flash biological effects has been established at Chang Gung Memorial Hospital. The final setup data represent a reference for the beam users to plan the experiments as well as for the improvement of the facility.

1. INTRODUCTION

The proton therapy center at Chang Gung Memorial Hospital in Taiwan started clinical operations in 2017. Currently, a cyclotron (230MeV, SHI, Japan) serves four medical treatment rooms, where more than 1500 patients have been treated annually. The proton center is also equipped with a research experimental room, as shown in figure 1(A). There are three beam directions after switching magnet for broad and small field proton irradiation and high energy neutron generation in the research room. The left beamline for broad field irradiating was equipped with a double scattering system to generate an irradiation field large enough to ensure the delivery of a dose with better than 90% uniformity over a circular area with a diameter of 10 centimeters. A wheel range modulator was installed additionally for spread-out Bragg peak (SOBP) in this line. Recently, ultra-high-dose-rate "FLASH" radiotherapy, has been among the hot topics in radiation oncology. Most experimental results have shown it has the healthy tissue sparing capability and may help in reducing collateral damage. These results, typically seen at dose rates exceeding 40 Gy/s irradiation delivered in short pulses,[1] have been widely reported in studies utilizing photon or electron radiation as well as in some proton radiation studies.[2–7] Among them, as protons have the characteristics of Bragg peak dose distribution, it is believed that proton Flash will be a powerful appliance for tumor treatment.[8,9]

Prior to clinical usage, it is necessary to understand the basic dosimetry related to biological effects at Flash condition.[10,11] Especially last year, the US FDA passed the registry of world's first proton flash human trial, which urges the proton Flash research[12]. Before Flash has been translated to clinical modality, not only the mechanism of biological effects need to be clarified, but also the related dosimetry protocols should to achieve consistency[11,13,14]. Due to the need for conducting investigations, we have constructed an irradiation platform at the 0 degrees direction beam line which the target dose rate can be delivered from conventional (~0.04 Gy/s) dose rates to Flash (>40 Gy/s) using scattered protons. Because the Flash effect has been observed for dose delivered time less than a second, it seems the underlying biological mechanism will be influenced on a comparable timescale[15]. Therefore, under this key limitation, a clear picture of the delivery time structure is critical for future studies on Flash biological effects. However, there are scarce commercially available dosimetry devices with precise time structure measurement for Flash conditions at present[16], so it is in need to develop reliable dose rate measuring apparatus with adequate temporal resolution[17,18]. The experimenters are able to distinguish that any observed biological effects arise not only the total dose but also from instantaneous dose rate delivered distinctions.

Here, we report the development of an experimental proton irradiation platform in a clinical proton facility as well as the measured results of the Flash beam characters. Irradiation procedure matched with an image guidance system to provide a precise proton irradiation for Flash biological studying. The established irradiation platform is not only suitable for Flash biological experiments but also for the development of related speedy proton beam measurement equipment. These data represent a reference for the beam users to plan the experiments as well as for the improvement of the facility.

## 2. MATERIALS AND METHODS

2.A General information about the irradiation platform

The beam produced from cyclotron, passes through transporting room into the research experimental room where the Flash irradiation platform was set up, as shown in figure 1(B). A 1 mm lead was used as a scatterer for beam flattening, and a set of two 60 mm thick copper bricks with circular holes were designed as collimators for shaping the scattered proton field size for the small animal. Although only pass-through irradiation experiments are currently performed, there is space for installing an attenuator for Bragg peak irradiation. The alignment between the two collimators is through a 2D array ion chamber (SINICA, Taiwan). There should be no physical connection between the research experiment room and the accelerator to avoid any possibility of interfering with clinical usage in which only the telephone is used as the beam delivery control method, as illustrated in figure 1. Due to the lack of a direct and rapid feedback mechanism for dose control, immediate monitoring of doses and dose rates in experiments is more critical.

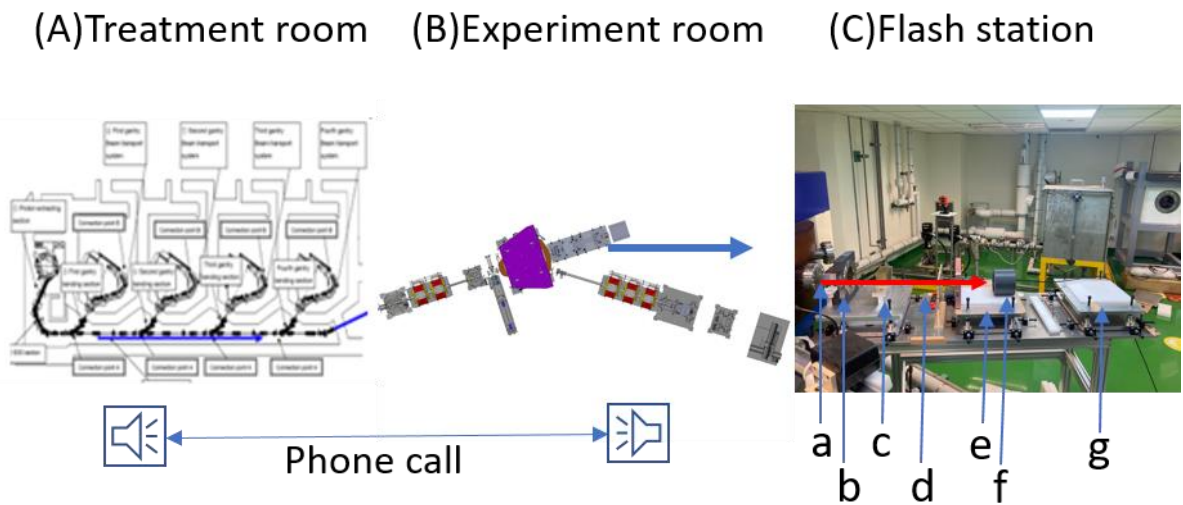

Figure 1. Configuration of Proton Center at Chang Gung Memorial Hospital, (A) cyclotron, treatment room and (B) experiment room (magnified), communication between cyclotron and experiment room by phone only (C) the picture of flash irradiation platform, where a is the collimator 1, b is the transmission ion chamber, c is the collimator 2, d is the sample position for pass through irradiating, e is the moveable table for range filter, f is the FC, g is the sample position for BP irradiation.

2.B Dose measurement devices

A 4 mm gap parallel electrodes transmission ionization chamber (TIC) with a 50 mm diameter 100 $\mu$m thick Kapton entrance window fill with one atm Ar gas was built as a beam dose/dose rate monitoring device. An 80 mm copper cylinder thicker than 230 MeV proton range was used as a Poor Man's Faraday Cup (PMFC) to collect absolute beam charge [19]. Both of them are equipped with a home-brewed fast reading current integral electronics device calibrated with a nano current source (Keithley Instruments, USA). An $1 \times 100 \times 100$ mm plastic scintillator with two contact image sensors along vertical and horizontal sides was used to be a beam position monitor (BPM) [20]. It was installed beyond TIC to monitor beam position and size. The schematic diagram of the PMFC, TIC, and BPM is shown in figure 2. At the target position, absolute dose measurements were performed with an Advanced Markus Chamber (PTW, Freiburg, Germany). Beam width, flatness, and symmetry were also evaluated at the target position using EBT3 films.

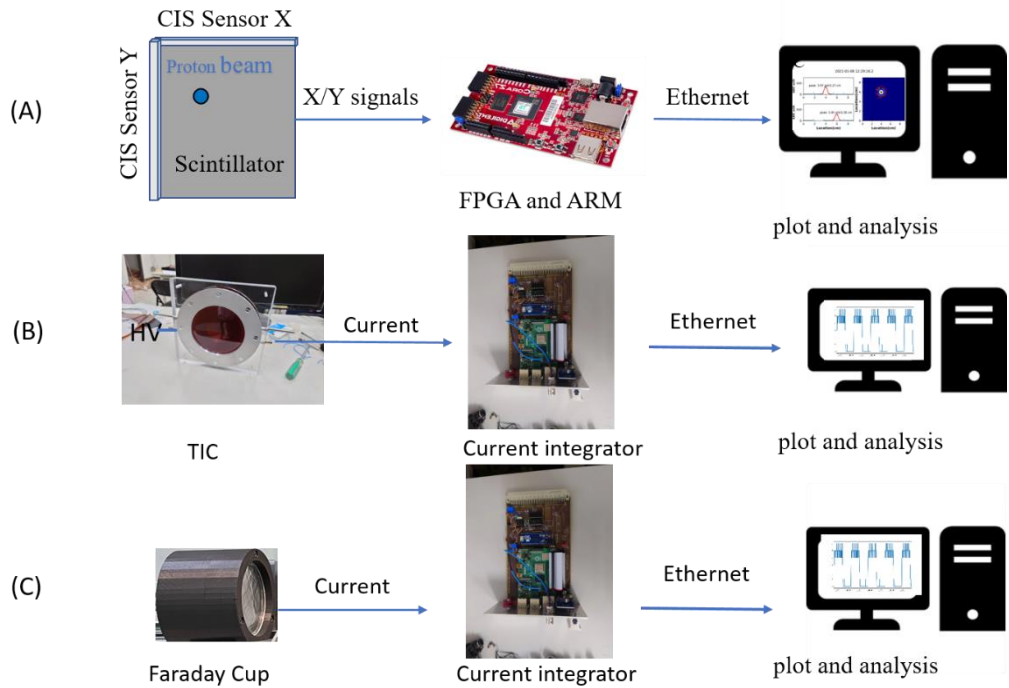

Figure 2. The schematic diagram of the (A) Scintillator based BPM, (B) TIC and (C) FC

2.C Small animal positioning

C57BL/6 mice were purchased from National Laboratory Animal Center (NLAC, Taipei, Taiwan) and the experimental protocol was approved by the Institutional Animal Care and Use Committee of Chang Gung University (IACUC：CGU105-067). The mice were initially anesthetized with Zoletil (40 mg/kg, Virbac, Pukete, Hamilton, New Zealand) and xylazine (20 mg/kg, Bayer Health Care Animal Health, Leverkusen, NRW, Germany) and then intravascularly injected with ExiTron nano 6000 contrast agent (Miltenyi Biotec, Bergisch Gladbach, DEU) for liver imaging. Mice were scanned by a moveable small animal computed tomography (You-Shang Technical CORP, TYN, TWN), and the images were analyzed by RadiAnt DICOM Viewer (Medixant, Poznán, POL) to visualize the location and morphology of liver tissues. An 1 mm-diameter solid-glass bead (Sigma-Aldrich, MO, USA) was used as the surface marker to mark the center of irradiation field. The irradiation field covered the median lobe of the liver.

2. D Immunohistological fluorescence staining

The γ-H2AX molecule is a common biomarker for labeling DNA double-strand breaks (DSB). To identify the region of proton irradiation, the liver tissues were stained by γ-H2AX antibody to examine the DNA damage. The mouse that received proton irradiation was sacrificed 30 minutes post-irradiation, and the irradiated liver tissue was imbedded by Tissue-Tek® O.C.T. compound. The 10 um tissue slides were fixed with cold methanol for 5 min and incubated with phosphate-buffered saline (PBS) with 1% bovine serum albumin (BSA) at room temperature for blocking the non-specific binding, and then anti-phospho-histone H2AX (γ-H2AX) (Cell Signaling Technology, MA, USA) at 4 °C overnight for DNA damage detection. After washing with PBS, the primary antibodies were detected by Alexa Fluor 488 goat anti-rabbit IgG (Invitrogen, CA, USA). The tissue slides were mounted with ProLong Gold Antifade Mountant with DAPI (Invitrogen). The images were captured by ImageXpress Micro Confocal High-Content Imaging System (Molecular Device, CA, USA).

3. RESULTS and DISCUSSION
3.A Beam intensity measurement

The transmission ionization chamber (TIC) is the main dose monitoring device. Ion recombination in ionization chambers is usually considered a side effect, which influences the accuracy of determining the absorbed dose[21,22]. Therefore a homemade TIC has to be verified its linearity with the intensity of the irradiating beam. The experimental configuration for testing the homemade TIC, FC, and BPM is shown in figure 3(A). Proton beam passed through TIC and BPM then stopped at FC. The measured results of one-shoot pulse are shown in figure 3(B), (C), and (D) correspondent to TIC, FC and BPM respectively. The linear regression of TIC and FC results under the limitation of cyclotron output is plotted on figure 3(E). The fitting result ($R^2$=0.99996) indicates that the ion recombination effect of the homemade TIC could be ignored under the testing beam intensity.

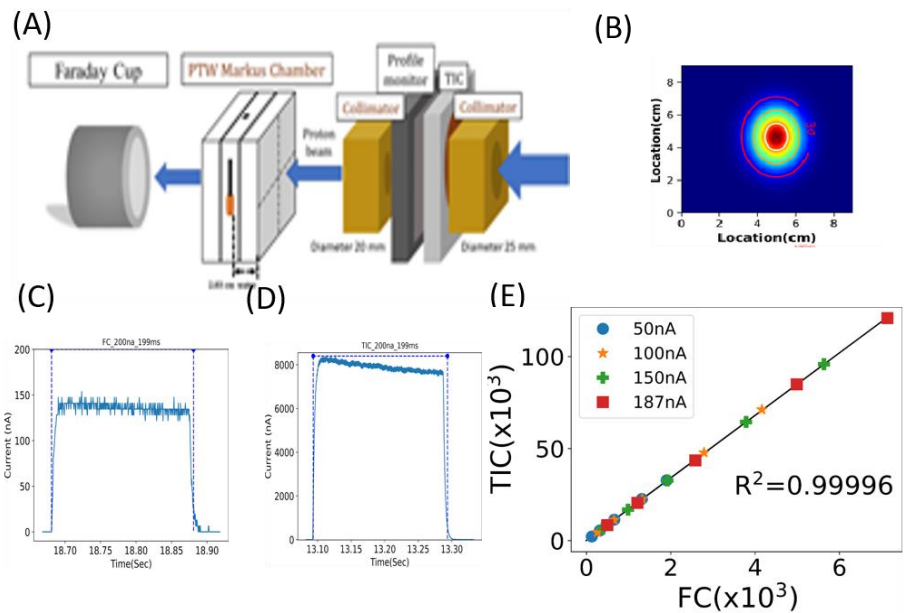

Figure 3. (A) Configuration of TIC, FC, and BPM tests, (B) the BPM result, (C), (D) the FC and TIC results, TIC v.s. FC with linear regression fit was shown on (E) where the symbols represent requested ion beam intensity.

Because the cyclotron provides only shorter than 200 mS pulse presently, the results of measuring different pulses with various pulse widths at fixed nominated beam intensity are shown in figure 4. The measured pulse duration are close to the setting time, but the measured beam current intensities are off the setting value pulse by pulse. The shortest pulse that could be detected was 1 ms. It implied that the developed integral charge measurement reading electronic was able to provide time structure measurement within 1 ms time resolution. To test the beam intensity stability, five continuous 199 ms pulse with 1 ms interruption was measured and shown in figure 5. The measured result revealed that the intensity was gradually reduced to 70% after the 5th pulse. This phenomenon seems to be caused by the cyclotron stabilizer not being activated in the short period of the flash pulse. In other words, the actual irradiating dose may be discrepant from the expected dose. It should be noted that when discussing Flash biological effects, there is no confusion due to the discrepancy of total doses between Flash and conventional irradiated treatment. In order to avoid the confusion caused by the discrepancy in total dose when comparing the biological effects between Flash and conventional irradiation head by head. It is therefore necessary to overcome the above possible problem before the beam intensity can be stably controlled in Flash irradiation. At this stage, the problem could be solved by the developed Flash dose monitoring device. Flash irradiation is performed first, and the total dose is confirmed by the result of the dose monitoring device. Then perform conventional irradiation at this dose to ensure that both doses are the same, only the dose rate is different. It is reasonable to prescript the conventional irradiation target total dose base on the real integral Flash pulse dose, not vice versa.

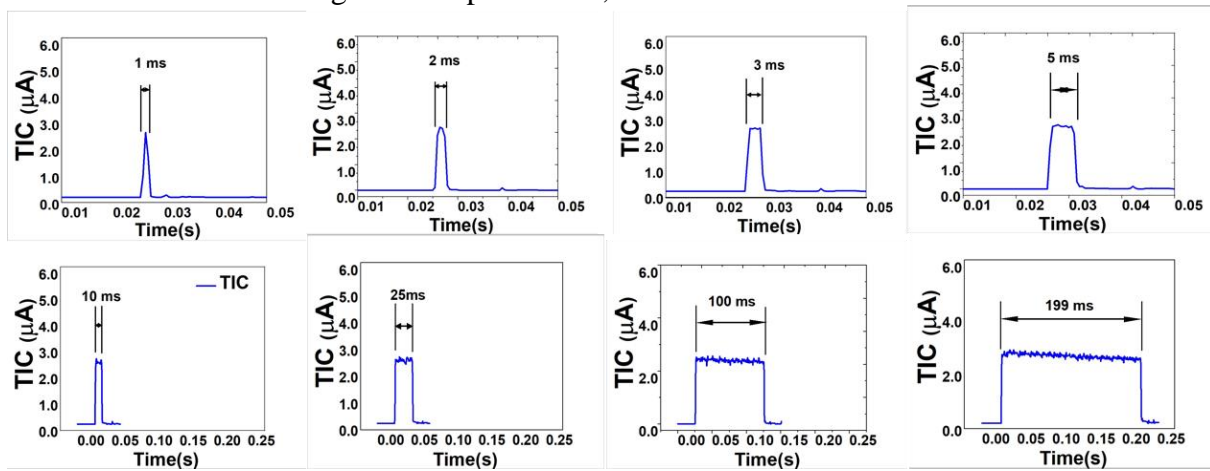

Figure 4. The TIC results of measuring pulses from 1 ms to 199 ms at a same requested beam intensity.

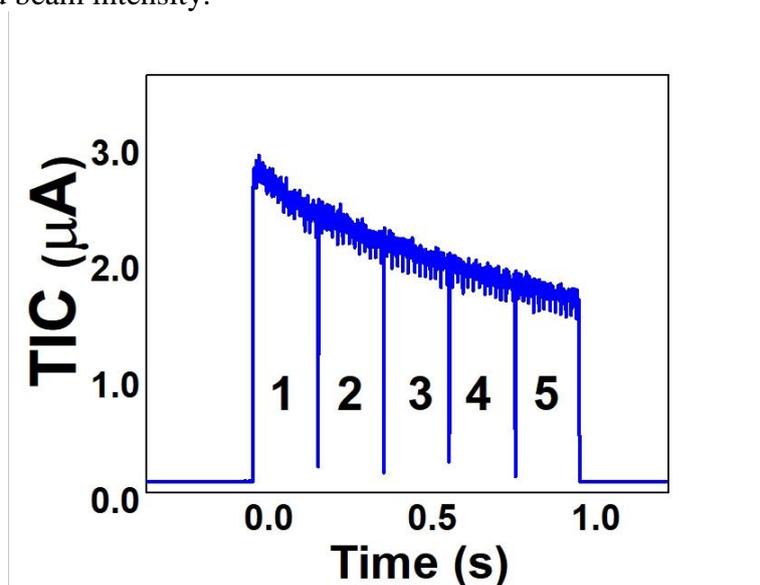

Figure 5 Five continuous 199 ms pulses with 1 ms interval off, measured by TIC.

## 3.B  Beam Profile and Absolute Dose Delivery

The irradiation area can be adjusted by the second collimator, namely the sample collimator. The dose homogeneity and absolute dose were investigated at the position of the small animal location using EBT3 films and the calibrated Advanced Markus Chamber. Figure 6(A) are the results of measuring the beam profiles of the small and large fields at the target position with EBT3 films. The resulting dose distributions revealed that the small irradiating field has a better homogeneity than the larger one. The measured dose distribution could be used as a reference for subsequent assessment of biological effects. The beam condition of a passive scatter system often depends on the geometric configuration, such as the thickness of scatterer, alignment of collimators, and the distance of target. For experiments requiring high beam uniformity, it might be achieved by adjusting those parameters. Besides, the plastic scintillator based BPM installed in the beamline could immediately provide beam position and size information as shown in figure 3(B). The linear relationship between the results of Advanced Markus Chamber and FC was shown in figure 6(B). The maximum dose rate measured by Advanced Markus Chamber was 106.15 Gy/s which corresponds to FC measured current 132.6 nA. The good linearity ($R^2=0.99996$) indicates that the Advanced Markus Chamber is suitable for measuring absolute dose under this high dose rate. Both the results of regression fitting of Advanced Markus Chamber VS FC (fig. 6(B)) and TIC VS FC (fig. 3(E)) show good linearity that confirmed the absolute dose of the irradiated object could be real-time monitored by the on-line installed TIC.

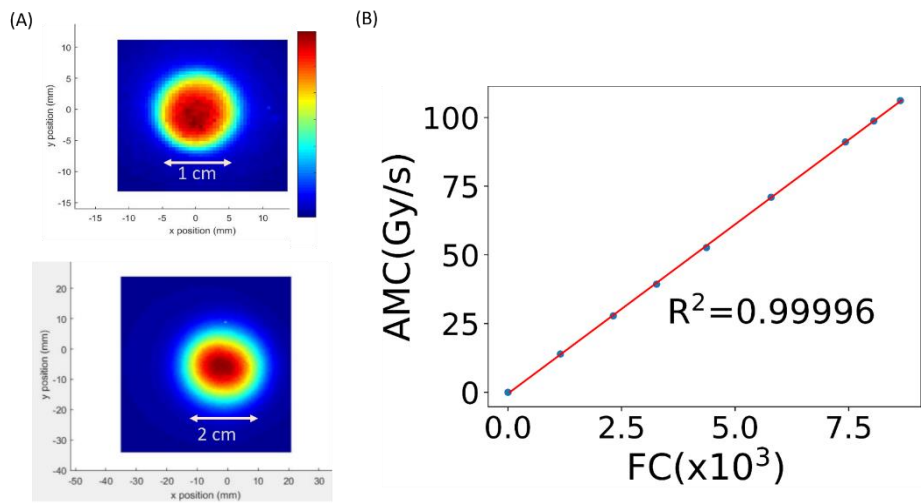

Figure 6. (A) EBT3 film results of the small and large field at the target position, (B) the results of regression fit of Advanced Markus Chamber VS FC

## 3.C  Image guidance and Experiment Procedure

The procedure of proton irradiation for small animals was shown in Figure 7. Mice were anesthetized and injected with ExiTron nano 6000 contrast agent. The region of liver tissue was shown by computed tomography scanning. Based on the result of the computed tomography scanning, a glass bead was attached to the body surface of mouse to be a mark of the center of irradiated field, as shown on the figure 7(A) to (C). Mouse with attached bead was transferred to the irradiation platform one by one, and aligned the bead with proton beam by a laser positioning device, as shown on figure 7(D). γ-H2AX, a common biomarker for labeling DNA DSB (double strand break), was utilized to identify the region of proton irradiated in this test experiment. A mouse was sacrificed 30 min after irradiation, and the damage region of liver tissues was examined by γ-H2AX staining. The DNA was located by DAPI staining. Un-irradiated liver tissue showed basal level of γ-H2AX signal. Irradiated liver tissues exhibited numerous γ-H2AX signal, it indicated high level of DSB occurred after proton irradiated, as shown on figure 7(E). These results demonstrated that the procedure of small animal irradiation guided by the imaging system could accurately perform localization irradiation on specific region *in vivo*. In practice, completing the steps for one sample takes about 10 minutes, including the small animal mount and waiting for the environment radiation level to be acceptable.

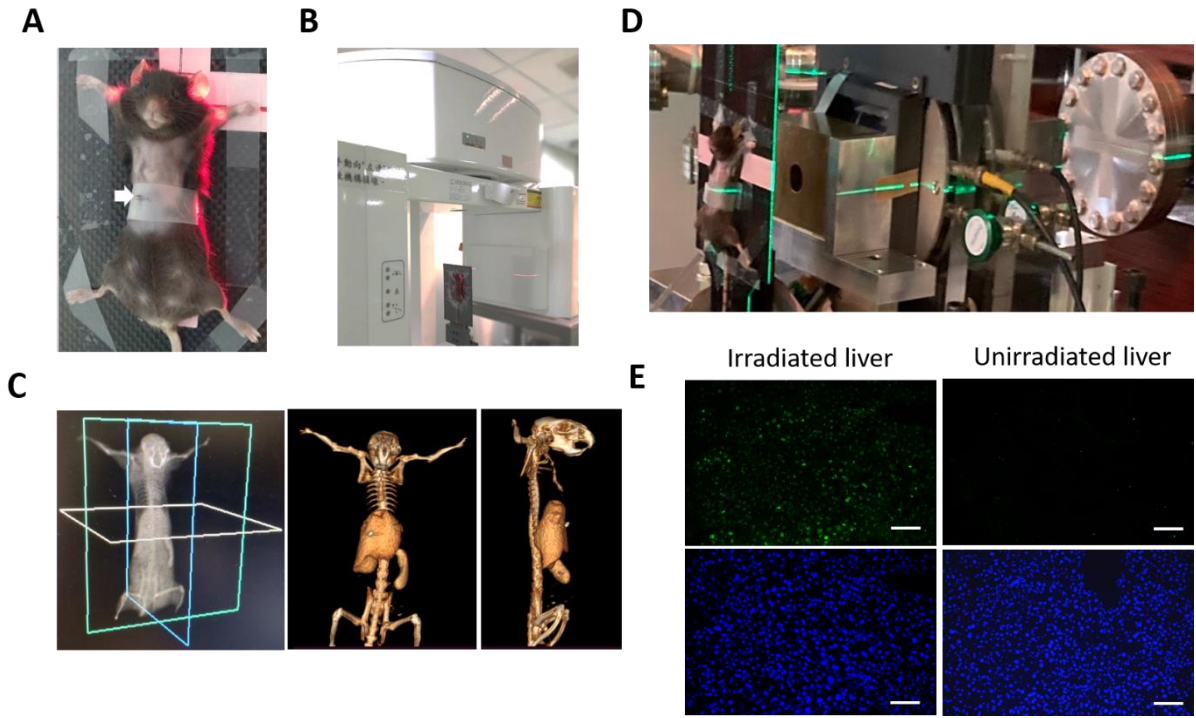

Figure 7. Proton irradiating procedure for small animal. (A) The anesthetized mouse mounted with a reference bead (arrow pointing). (B) CT scanning, (C) 3D image, (D) Laser positioning with the reference bead, (E) Upper (green) and bottom (blue) pictures with respect to γ-H2AX and DAPI stain results, up-left picture shows high level γ-H2AX signal than up-right one.

4. Conclusion

A proton research irradiation platform dedicated for studying small animal Flash biological effects has been established at Chang Gung Memorial Hospital. The flexible design of the irradiation platform can be adjusted by the experimenters according to their needs. The ultimate dose and dose rate at present could reach about 20 Gy and 100 Gy/s, respectively. A homemade TIC was equipped to monitor irradiation beam intensity with a time structure that immediately provided both the total dose and the variation of dose rate. Besides, with the CT image guidance system, the delivered dose of the small animal irradiated region can be accurately evaluated. Animal irradiation experiments being in progress, the results will be published separately when they are carried out.

Acknowledgements

This work was partially financial supported by NSTC-110-2221-E-007-024 and supported in part financially by Chang Gung Memorial Hospital, Linkou Research Grants CMRPG3K1911, CMRPG3L1381, and CMRPG3M0681. We also thank the Particle Physics and Beam Delivery Core Laboratory, Chang Gung Memorial Hospital, Linkou for support of the experiment.

REFERENCES

1. Bourhis J, Montay-Gruel P, Gonçalves Jorge P, et al. Clinical translation of FLASH radiotherapy: Why and how? *Radiother Oncol*. 2019;139:11-17. doi:10.1016/j.radonc.2019.04.008

2. Montay-Gruel P, Bouchet A, Jaccard M, et al. X-rays can trigger the FLASH effect: Ultra-high dose-rate synchrotron light source prevents normal brain injury after whole brain irradiation in mice. *Radiother Oncol*. 2018;129(3):582-588. doi:10.1016/j.radonc.2018.08.016

3. Montay-Gruel P, Corde S, Laissue JA, Bazalova-Carter M. FLASH radiotherapy with photon beams. *Med Phys*. 2022;49(3):2055-2067. doi:10.1002/mp.15222

4. Jaccard M, Durán MT, Petersson K, et al. High dose-per-pulse electron beam dosimetry: Commissioning of the Oriatron eRT6 prototype linear accelerator for preclinical use. *Med Phys*. 2018;45(2):863-874. doi:10.1002/mp.12713

5. Schüler E, Acharya M, Montay-Gruel P, Loo BW, Vozenin M, Maxim PG. Ultra-high dose rate electron beams and the FLASH effect: From preclinical evidence to a new radiotherapy paradigm. *Med Phys*. 2022;49(3):2082-2095. doi:10.1002/mp.15442

6. Diffenderfer ES, Sørensen BS, Mazal A, Carlson DJ. The current status of preclinical proton FLASH radiation and future directions. *Med Phys*. 2022;49(3):2039-2054. doi:10.1002/mp.15276

7. Sørensen BS, Sitarz MK, Ankjærgaard C, et al. In vivo validation and tissue sparing factor for acute damage of pencil beam scanning proton FLASH. *Radiother Oncol*. 2022;167:109-115. doi:10.1016/j.radonc.2021.12.022

8. Kourkafas G, Bundesmann J, Fanselow T, et al. FLASH proton irradiation setup with a modulator wheel for a single mouse eye. *Med Phys*. 2021;48(4):1839-1845. doi:10.1002/mp.14730

9. Kim MM, Verginadis II, Goia D, et al. Comparison of FLASH Proton Entrance and the Spread-Out Bragg Peak Dose Regions in the Sparing of Mouse Intestinal Crypts and in a Pancreatic Tumor Model. *Cancers*. 2021;13(16):4244. doi:10.3390/cancers13164244

10. Coleman CN, Ahmed MM. Implementation of New Biology-Based Radiation Therapy Technology: When Is It Ready So "Perfect Makes Practice?" *Int J Radiat Oncol Biol Phys*. 2019;105(5):934-937. doi:10.1016/j.ijrobp.2019.08.013

11. Guerrieri P, Jacob NK, Maxim PG, et al. Three discipline collaborative radiation therapy (3DCRT) special debate: FLASH radiotherapy needs ongoing basic and animal research before implementing it to a large clinical scale. *J Appl Clin Med Phys*. 2022;23(4). doi:10.1002/acm2.13547

12. Breneman J. *Feasibility Study of FLASH Radiotherapy for the Treatment of Symptomatic Bone Metastases*. clinicaltrials.gov; 2021. Accessed June 15, 2022. https://clinicaltrials.gov/ct2/show/NCT04592887

13. Vozenin MC, Montay-Gruel P, Limoli C, Germond JF. All Irradiations that are Ultra-High Dose Rate may not be FLASH: The Critical Importance of Beam Parameter Characterization and In Vivo Validation of the FLASH Effect. *Radiat Res*. 2020;194(6):571-572. doi:10.1667/RADE-20-00141.1

14. Taylor PA, Moran JM, Jaffray DA, Buchsbaum JC. A roadmap to clinical trials for FLASH. *Med Phys*. 2022;49(6):4099-4108. doi:10.1002/mp.15623

15. Kacem H, Almeida A, Cherbuin N, Vozenin MC. Understanding the FLASH effect to unravel the potential of ultra-high dose rate irradiation. *Int J Radiat Biol*. 2022;98(3):506-516. doi:10.1080/09553002.2021.2004328

16. Yang Y, Shi C, Chen C, et al. A 2D strip ionization chamber array with high spatiotemporal resolution for proton pencil beam scanning FLASH radiotherapy. *Med Phys*. Published online May 30, 2022:mp.15706. doi:10.1002/mp.15706

17. El Naqa I, Pogue BW, Zhang R, Oraiqat I, Parodi K. Image guidance for FLASH radiotherapy. *Med Phys*. 2022;49(6):4109-4122. doi:10.1002/mp.15662

18. Karsch L, Pawelke J, Brand M, et al. Beam pulse structure and dose rate as determinants for the flash effect observed in zebrafish embryo. *Radiother Oncol*. 2022;173:49-54. doi:10.1016/j.radonc.2022.05.025

19. Cascio EW, Gottschalk B. A Simplified Vacuumless Faraday Cup for the Experimental Beamline at the Francis H. Burr Proton Therapy Center. In: *2009 IEEE Radiation Effects Data Workshop*. ; 2009:161-165. doi:10.1109/REDW.2009.5336294

20. Hsiao TY, Niu H, Chen TY, Chen CH. Develop a high energy proton beam position monitor using linear contact image sensor. *MethodsX*. 2020;7:100773. doi:10.1016/j.mex.2019.100773

21. Kranzer R, Poppinga D, Weidner J, et al. Ion collection efficiency of ionization chambers in ultra-high dose-per-pulse electron beams. *Med Phys*. 2021;48(2):819-830. doi:10.1002/mp.14620

22. Mirandola A, Maestri D, Magro G, et al. Determination of ion recombination and polarity effects for the PTW Advanced Markus ionization chamber in synchrotron based scanned proton and carbon ion beams. *Phys Med*. 2022;96:149-156. doi:10.1016/j.ejmp.2022.03.007